\documentclass[doublecol]{epl2} 
\usepackage{amssymb}
\usepackage{amsmath}
\usepackage[normalem]{ulem}

\newcommand{\bs}[1]{{\boldsymbol{#1}}}
\newcommand{\bk}{\bs{k}}
\newcommand{\br}{\bs{r}}

\title{Weakly interacting disordered Bose gases out of equilibrium: from multiple scattering to superfluidity}
\shorttitle{Weakly interacting disordered Bose gases out of equilibrium} 

\author{T. Scoquart\inst{1} \and  P.-\'E. Larr\'{e}\inst{2} \and D. Delande\inst{1} \and N. Cherroret\inst{1}}
\shortauthor{T. Scoquart \etal}

\institute{                    
  \inst{1} Laboratoire Kastler Brossel, Sorbonne Universit\'e, CNRS, ENS-Universit\'e PSL, Coll\`ege de France; 4 Place Jussieu, 75005 Paris, France\\
  \inst{2} CY Cergy Paris Universit\'e, CNRS, LPTM, F-95000 Cergy, France
}

\abstract{
We explore the quench dynamics of a two-dimensional, weakly interacting disordered Bose gas for various relative strengths of interactions and disorder. This allows us to identify two well distinct out-of-equilibrium regimes. 
When interactions are smaller than the disorder,  the gas experiences multiple scattering and exhibits a short-range spatial coherence. At short time this coherence is only smoothly affected by interactions, via a diffusion process of the particles' energies. 
When interactions are larger than the disorder, scattering ceases and the gas behaves more and more like a fluid, ultimately like a superfluid at low energy. In the superfluid regime, the gas exhibits a long-range algebraic coherence, characteristic of a pre-thermal regime in disorder.
}

\begin{document}

\maketitle

\section{Introduction}

In quantum gases, the interplay between disorder and interactions leads to a rich variety of phenomena. At low temperature for instance, a one-dimensional Bose gas in equilibrium  behaves either as an  insulator or a superfluid  depending on the relative strength of disorder and interactions \cite{Cazalilla11}. The insulating phase has been predicted to be robust at finite temperature, up to a critical point known as the many-body localization transition \cite{Aleiner10}. Initially described for interacting  electrons in dirty conductors \cite{Gornyi05, Basko06}, many-body localization  has, since then, triggered a considerable interest in various fields of physics \cite{Abanin19}. 
Another intriguing problem is the dynamical evolution of interacting disordered gases brought \textit{out-of-equilibrium} by an initial quench. In the homogeneous case, out-of-equilibrium interacting gases generically thermalize to a universal Gibbs ensemble \cite{Polkovnikov11}. For sufficiently strong disorder and interactions on the other hand, many-body localization may prevent thermalization to occur \cite{Abanin19}. In between these two limits, the quench dynamics of disordered quantum gases remains  largely unexplored, in particular in dimensions larger than one.


In this context, a yet poorly explored problem is the out-of-equilibrium dynamics of \textit{weakly interacting} disordered gases. In this regime, mostly states in the ergodic phase of the many-body-localization transition contribute to the dynamics, so that thermalization is the rule. For bosons, the weakly interacting regime can be tackled at the mean-field level using the disordered Gross-Pitaevskii equation. The dynamical establishment of thermalization in this context was  previously studied in one-dimensional disordered chains \cite{Basko11, Kottos11} and in two-dimensional (2D) \cite{Cherroret15, Scoquart2020} random potentials, in cases where the disorder is typically \textit{larger} than interactions. In this regime, the short-time dynamics of the gas is governed by multiple scattering, while interactions make the system thermalize on a longer time scale. The opposite limit of a disorder \textit{smaller} than interactions has, on the other hand, little been explored in an out-of-equilibrium context. 

In this Letter, we study the quench dynamics of a 2D weakly interacting disordered Bose gas by varying the relative strengths of disorder and interactions over a wide range. We  characterize this dynamics by numerically and analytically computing the momentum distribution and the coherence function of the gas, focusing on relatively short times, before the system is fully thermalized. For a disorder larger than interactions, we recover  that the gas experiences multiple scattering, while being slowly thermalized by interactions. The spatial coherence is low in this limit. Upon increasing the strength of interactions, we find that scattering diminishes and that the gas behaves more and more like a fluid. 
Ultimately, when interactions become larger than the disorder, the gas becomes a non-equilibrium disordered superfluid with a long-range, algebraic coherence emerging in a light-cone fashion, which is typical of a 2D pre-thermalization process.

\section{The model}

Following the approach of \cite{Cherroret12, Muller12, Labeyrie12, Muller15, Ghosh14, Cherroret15 ,Scoquart2020}, we consider a simple out-of-equilibrium protocol where a 2D $N$-particle Bose gas, initially prepared in the plane-wave state $|\bk_0\rangle$, undergoes at $t=0$ a simultaneous interaction and disorder-potential quench. We describe the ensuing dynamics for $t>0$ with the time-dependent disordered Gross-Pitaevskii equation
\begin{equation}
i\partial_t\Psi(\br,t)= \left[-\frac{\boldsymbol{\nabla}^2}{2m} + V(\br)+gN|\Psi(\br,t)|^2\right]
\Psi(\br,t)
\label{eq:grosspitaevskii}
\end{equation}
for the Bose mean field $\Psi(\br,t)$, normalized according to $\int d^2\br |\Psi(\br,t)|^2=1$. Here and in the rest of the Letter, we set $\hbar=1$. We choose the random potential $V(\textbf{r})$ to be Gaussian distributed, with mean value $\overline{V(\textbf{r})}=0$ and correlation function $\overline{V(\textbf{r})V(\textbf{r}')}=V_0^2\exp(-|\br-\br'|^2/2\sigma^2)$, where $V_0$ is the amplitude of disorder fluctuations, $\sigma$ the correlation length, and the overbar refers to disorder averaging. In our numerical simulations, we generate this potential in a standard way, by convoluting a spatially $\delta-$correlated Gaussian random field with a Gaussian cutoff function \cite{Huntley89, Horak98}.
The temporal propagation of the initial plane wave is performed using a split-step algorithm that includes a Chebyshev expansion of the linear part of the evolution operator, as explained in \cite{Scoquart2020}. Simulations are performed on a 2D regular grid of size $L\times L$ with periodic boundary conditions along $x$ and $y$. A cell of surface $(\pi\sigma)^2$ is discretized in typically 5 to 7 steps along both $x$ and $y$. Throughout the paper, numerical values of lengths, momenta, energies and times
will be given in units of $\sigma$, $\sigma^{-1}$, $1/(m\sigma^2)$ and $m\sigma^2$, respectively. Finally, all the results are averaged over 3200 to 4800 disorder realizations.

\section{Momentum distributions}

\begin{figure*}[h]
\onefigure[scale=0.24]{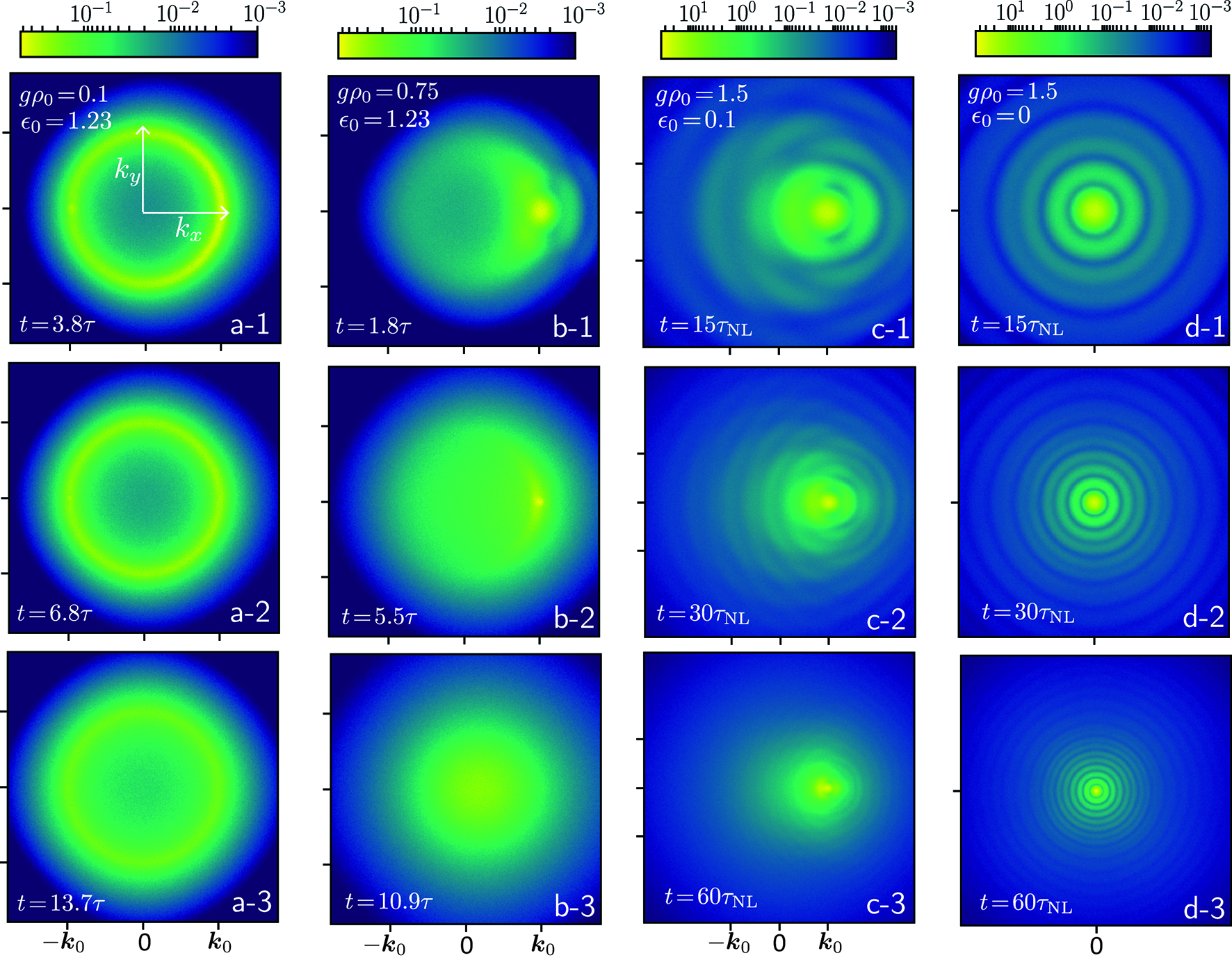}
\caption{Momentum distribution $n_\bk(t)$ of an interacting Bose gas launched with momentum $\bk_0$ in a 2D random potential.
Rows correspond to three successive times, given either in units of the scattering mean free time $\tau$ or of the nonlinear time $\tau_\text{NL}$. For all plots the disorder amplitude is set to $V_0=0.4$. 
(a) $\epsilon_0>V_0>g\rho_0$: particles experience multiple scattering on the random potential (with here $k_0\ell=18.0$). This gives rise to an incoherent elastic ring in momentum space. As time increases, the distribution slowly thermalizes due to particle  collisions. 
(b) $\epsilon_0> g\rho_0> V_0$: increased interactions compete with disorder and thermalize the distribution before the elastic ring is fully formed.
(c) $g\rho_0> \epsilon_0> V_0$: interactions are larger than the kinetic energy, so that the gas starts to behave as a superfluid. The initial small disorder-induced field fluctuations are coherently enhanced and yield an interference ring pattern in momentum space.
(d) $g\rho_0> V_0>\epsilon_0=0$: superfluid regime for a gas initially at rest.
}
\label{fig_1}
\end{figure*}

We first  compute numerically the disorder-averaged momentum distribution $n_\bk(t)\equiv \overline{|\langle\bk|\Psi(t)\rangle|^2}$ from Eq. (\ref{eq:grosspitaevskii}) using the procedure explained above, for various relative values of the disorder potential fluctuations, $V_0$, initial kinetic energy $\epsilon_0\equiv k_0^2/2m$ and initial interaction energy $g\rho_0\equiv gN/V$, where $V=L^2$ is the volume of the system. The distribution is normalized according to $\int d^2\bk/(2\pi)^2n_\bk(t)=1$. 
Density plots of $n_\bk(t)$ are shown in Fig. \ref{fig_1}, with the columns corresponding to different parameter regimes and the rows to three successive times, chosen relatively short compared to the time where the gas gets completely thermalized by interactions.\newline

$\bullet$ Regime $\epsilon_0>V_0> g\rho_0$ (Figs. \ref{fig_1}a): Particles of energy $\sim\epsilon_0$ experience elastic multiple scattering on the disorder potential, which randomizes the direction of their momenta. The distribution $n_\bk(t)$ thus quickly becomes a ring of radius $k_0$. As time increases (from Fig. \ref{fig_1}a-1 to a-3), the ring is slowly smoothed by particle collisions (thermalization). \newline

$\bullet$ Regime $\epsilon_0> g\rho_0>V_0$ (Figs. \ref{fig_1}b):
When the interaction strength is increased, particle collisions scramble the distribution before the scattering ring is fully formed. \newline

$\bullet$ Regime $g\rho_0>\epsilon_0> V_0$ (Figs. \ref{fig_1}c):
When interactions become larger than the initial kinetic energy, the gas starts to behave as a superfluid. There is no longer scattering, rather the disorder-induced field fluctuations are coherently enhanced and manifest themselves as interference rings, which become tighter and tighter as time increases. \newline

$\bullet$ Regime $g\rho_0> V_0>\epsilon_0$ (Figs. \ref{fig_1}d):
Same as in Figs. \ref{fig_1}c, but for a Bose gas initially almost motionless, at rest on the plots ($\epsilon_0=0$).\\

To better understand the physics at play in the distributions of Fig. \ref{fig_1}, we now discuss more in details the two extreme regimes of Figs. \ref{fig_1}a and \ref{fig_1}d.

\section{Multiple-scattering regime}

\begin{figure}
\onefigure[scale=0.35]{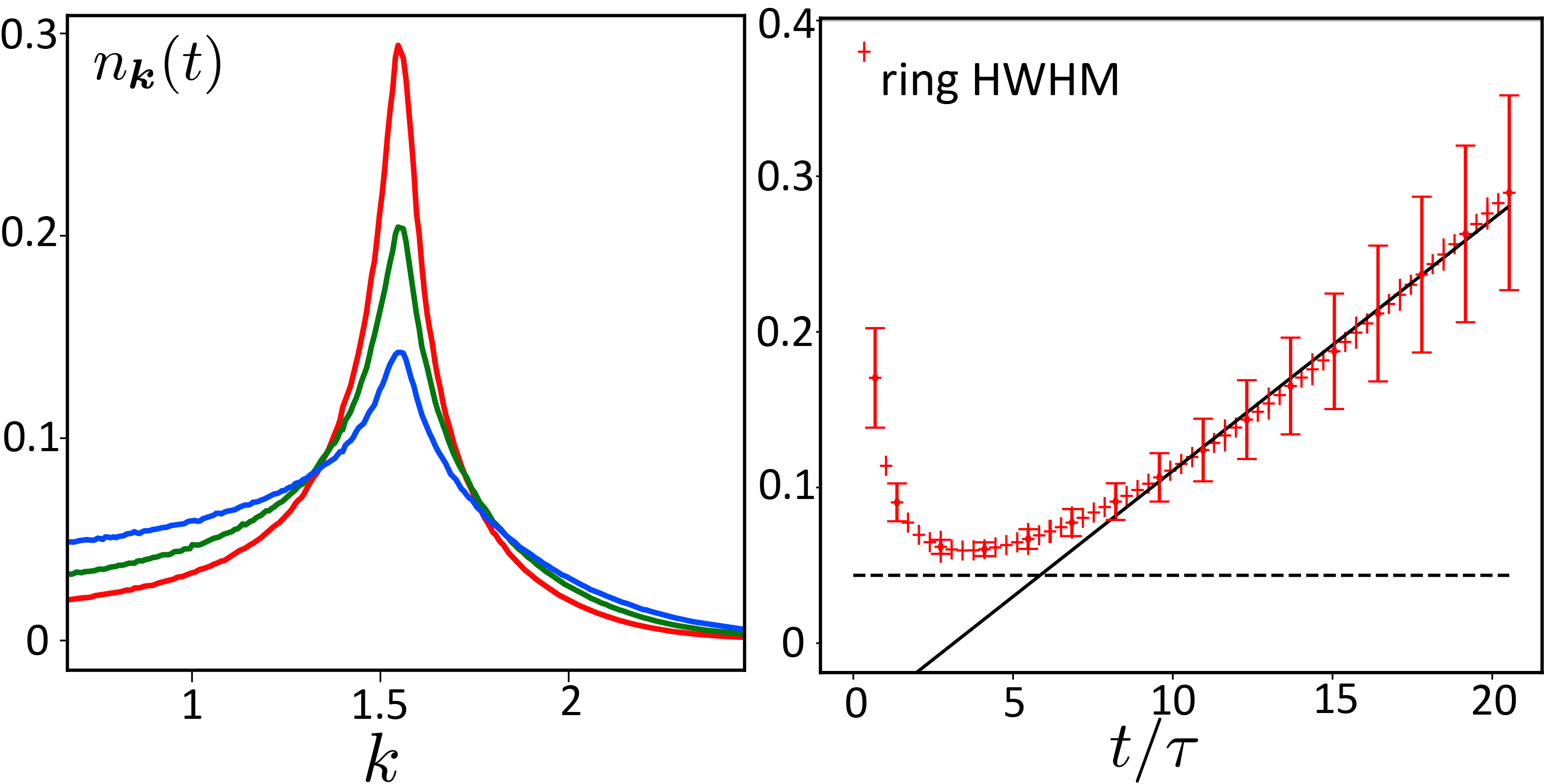}
\caption{
\label{ring_width_fig}
Left: radial cuts of the scattering ring, computed numerically at times $t=10.9\tau$, $15.1\tau$ and $19.8\tau$ from top to bottom (numerical data points have been joined for clarity). Each cut involves an angular average of the momentum distribution.
Right: half width at half maximum (HWHM) of the ring,  obtained by fitting the radial cuts with a Lorentzian profile in a $k$-window of width $0.8$ centered on $k_0$ (red symbols). Error bars are deduced from the $\chi^2$ goodness of the fits.
The solid line is a linear fit $\Delta k_\text{fit}(t)=1/\ell[1+6.38(t-5.84\tau)/\tau_\text{col}]$, which confirms the theoretical expectation. The dashed line shows the $g=0$ result, $\Delta k=1/\ell$.
Here $g\rho_0=0.07$, $V_0=0.2$, $k_0=1.57$, $k_0\ell=36$ and the chosen system size is $L=200\pi$.  With these parameters, $\tau=14.6$ and $\tau_\text{col}=251.5$.}
\end{figure}

We first address the low-interaction regime of Figs. \ref{fig_1}a. Since interactions are typically the smallest energy scale in this limit, the physics at short time is essentially a multiple-scattering process of quasi-particles with energy $\epsilon_k\equiv \bk^2/2m$ in the random potential. The relevant time scale is here the scattering mean free time $\tau$, which gives the rate at which these quasi-particles are elastically scattered. For the chosen Gaussian potential, we have
\begin{equation}
\frac{1}{\tau}=2\pi V_0^2\sigma^2e^{-k_0^2\sigma^2}I_0(k_0^2\sigma^2)
\end{equation}
in the Born approximation \cite{Akkermans07}, where $I_0$ is the modified Bessel function of the first kind. When $t\gg\tau$, the disorder-averaged momentum distribution acquires a ring shape, as seen in Figs. \ref{fig_1}a. For weak interactions, it was shown to be given by \cite{Scoquart2020}
\begin{equation}
n_\bk(t)=\int d\epsilon A_\epsilon(\bk)f_\epsilon(t)/\nu,
\label{nbk_largeg_eq}
\end{equation}
where $f_\epsilon(t)$ is the energy distribution of the quasi-particles (normalized according to $\int d\epsilon f_\epsilon(t)=1$), $A_\epsilon(\bk)$ is their spectral function 
and $\nu=m/2\pi$ is the 2D density of states per unit volume.
In the weak-disorder regime where $k_0\ell\gg1$, we have $A_\epsilon(\bk)\simeq1/(2\pi\tau)\times1/[(\epsilon-\epsilon_k)^2+1/4\tau^2]$. In the absence of interactions, $f_\epsilon(t)=A_\epsilon(\bk_0)$ is independent of time and coincides  with the spectral function at $\bk=\bk_0$. Equation (\ref{nbk_largeg_eq}) then reduces to \cite{Cherroret12}
\begin{equation}
n_\bk(g=0)=\frac{1}{\pi\nu\tau}\frac{1}{(\epsilon_k-\epsilon_0)^2+1/\tau^2},
\label{ring_profile}
\end{equation}
which indeed describes a ring of half-width at half maximum $\Delta k\sim 1/\ell$, where $\ell=k_0\tau/m$ is the mean free path.
When $g\ne 0$, the energy distribution $f_\epsilon(t)$ evolves in time due to particle collisions. 
At weak disorder, its dynamics is governed by a Boltzmann-like kinetic equation \cite{Schwiete13, Cherroret15, Scoquart2020}. The particle collisions  occur at a mean rate given by the Fermi golden rule 
\begin{equation}
\frac{1}{\tau_\text{col}}\!=\!2\pi\int\!\!\frac{d^2\bk}{(2\pi)^2}
\overline{|\langle\bk_0|gN|\Psi|^2|\bk\rangle|^2}
\delta(\epsilon_0-\epsilon_\bk)
\!\sim\!\frac{(g\rho_0)^2}{\epsilon_0}.
\label{eq:self_e_born}
\end{equation}
While $f_\epsilon(t)$ ultimately becomes thermal when $t\gg\tau_\text{col}$ \cite{Cherroret15}, as long as $t<\tau_\text{col}$ the scattering ring remains well resolved  (as in Figs. \ref{fig_1}a-1-3) and we have found that the time evolution of $f_\epsilon(t)$ is approximately captured by a diffusion process in the energy domain:
\begin{equation}
f_\epsilon(t)\simeq \int d\epsilon' \frac{A_{\epsilon'}(\bk_0)}{\sqrt{4\pi Dt}}
\exp\left[-\frac{(\epsilon-\epsilon')^2}{4Dt}\right]
\label{gaussian_law_eq}
\end{equation}
with a diffusion coefficient in energy space such that $2D\tau_\text{col}\sim 1/\tau^2$. When $g\ne 0$, the momentum distribution at short time is thus obtained by convoluting the non-interacting result (\ref{ring_profile}) with the Gaussian distribution in Eq. (\ref{gaussian_law_eq}). For $\tau\ll t< \tau_\text{col}$ this leads to a broadening of the ring with half width $\Delta k(t)\sim (1+C t/\tau_\text{col})/\ell$, with $C$ a numerical constant. 
This broadening is visible in the left panel of Fig. \ref{ring_width_fig}, which shows radial cuts of the scattering ring, obtained numerically at successive times. As long as  $t<\tau_\text{coll}$ we find that the widths extracted from these cuts indeed increase linearly in time, as shown in the right panel of Fig. \ref{ring_width_fig}.

\begin{figure}
\onefigure[scale=0.42]{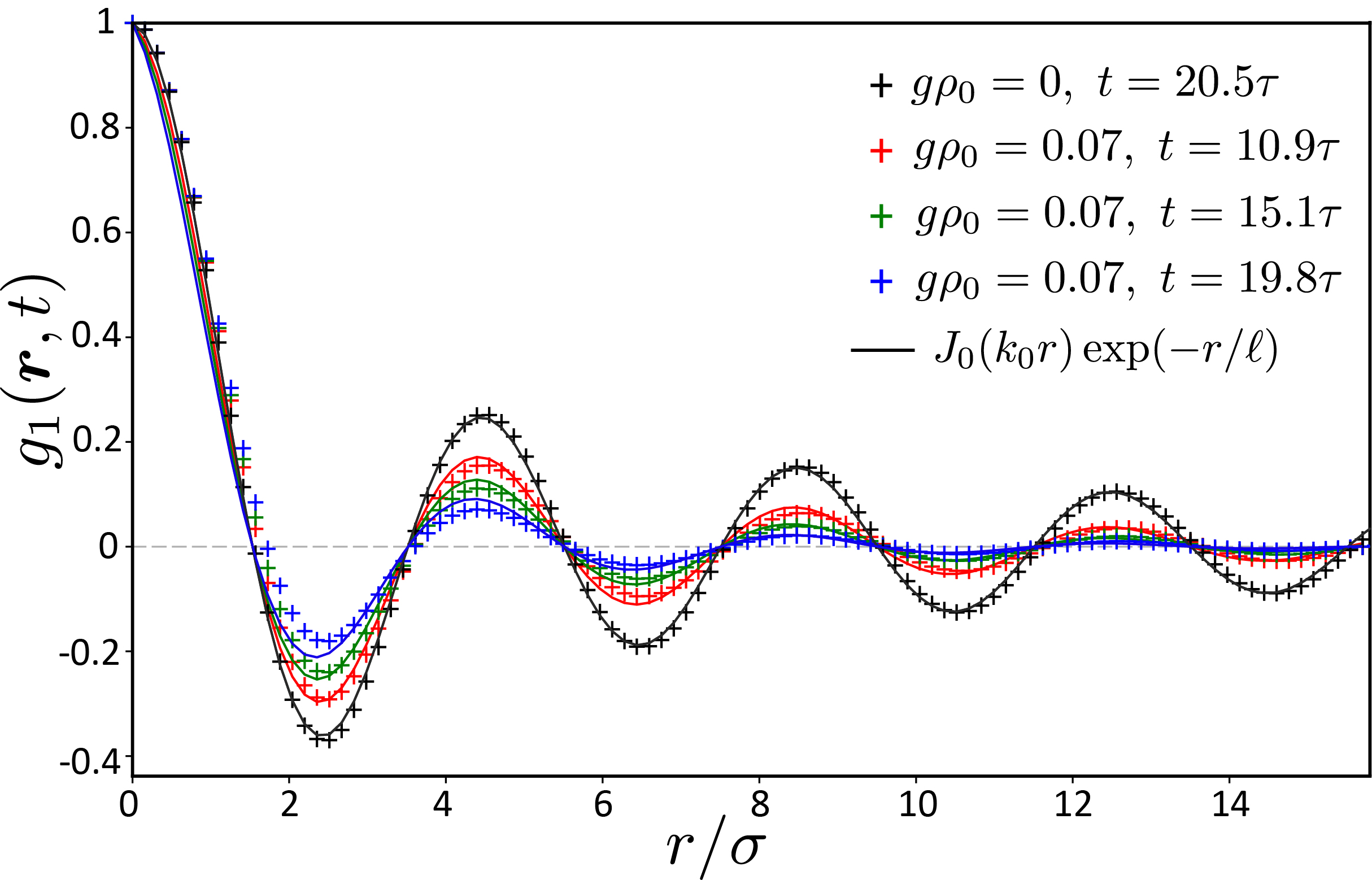}
\caption{
\label{g1_smallg_fig}
Radial cut of the coherence function $g_1(\br,t)$ of the Bose gas in the multiple scattering regime \cite{footnote1} (the cut also involves an angular average of $g_1$). For $g=0$ (outer black symbols, numerics, and solid black curve, theory) the function is independent of time after a few $\tau$ and exhibits oscillations at the scale $2\pi/k_0$. In the presence of interactions, these oscillations are smoothed. 
Colored solid curves for $g\ne 0$ are the theoretical prediction, Eq. (\ref{g1_smallg_theory}), in which we take the value $\Delta k(t)=\Delta k_\text{fit}(t)$ given in the caption of Fig. \ref{ring_width_fig}. 
Parameters are the same as in Fig. \ref{ring_width_fig}.
}
\end{figure}

It is also interesting to investigate the spatial coherence of the Bose gas in the multiple-scattering regime. To this aim, we consider the coherence function
 \begin{equation}
 g_1(\br,t)\equiv\overline{\Psi^*(0,t)\Psi(\br,t)}=\!\int\!\!\frac{d^2\bk}{(2\pi)^2}n_\bk(t)e^{i\bk\cdot \br}.
 \end{equation}
When $g=0$, $n_\bk$ is given by Eq. (\ref{ring_profile}), so that $g_1(\br,t)=J_0(k_0 r)\exp(-{r/\ell})$, with $r\equiv |\br|$ and $J_0$ the Bessel function of the first kind \cite{Akkermans07}. 
This expression is essentially the first-order correlation function of the 2D matter-wave speckle pattern formed by the particles scattered on the random potential \cite{Cherroret08}. It shows that the coherence length of the Bose gas is rather short in the multiple-scattering regime, of the order of the de Broglie wavelength $2\pi/k_0$. We show in Fig. \ref{g1_smallg_fig} the $ g_1$ function computed numerically from Eq. (\ref{eq:grosspitaevskii}) for $g=0$, together with the above theoretical formula. In the presence of interactions and as long as $t<\tau_\text{col}$, the shape of $n_\bk$ remains  close to the Lorentzian (\ref{ring_profile}), with a broadened width $\Delta k(t)\sim(1+Ct/\tau_\text{col})/\ell$, such that:
 \begin{equation}
 g_1(\br,t)\simeq J_0(k_0r)\exp[-\Delta k(t)r].
 \label{g1_smallg_theory}
 \end{equation}
As shown in Fig. \ref{g1_smallg_fig},  Eq. (\ref{g1_smallg_theory}) describes very well the numerical results for $g_1$ at $g\ne 0$. Here the main effect of interactions is  to smooth the matter-wave speckle, without significantly affecting the coherence of the gas. This picture would of course change at  long time $t\gg \tau_\text{col}$, where the momentum distribution starts to significantly deviate from Eq. (\ref{ring_profile}) and approaches a thermal law \cite{Cherroret15}. 

\section{Superfluid regime}

We now address the opposite, superfluid regime where $g\rho_0\gg V_0\gg \epsilon_0$. From now on, we focus for simplicity on the case $\epsilon_0=0$ of a gas  initially at rest, corresponding to the numerical distributions in Figs. \ref{fig_1}d-1-3. 
In such a low-energy regime, the natural approach for describing the dynamics of the 2D Bose gas relies on the density-phase representation of the wavefunction \cite{Pitaevskii16, Castin03}:
\begin{equation}
\Psi(\br,t)=\sqrt{\rho(\br,t)}\exp[i\theta(\br,t)-ig\rho_0 t],
\end{equation}
where $-ig\rho_0 t$ is the dynamical phase of the uniform solution in the absence of disorder ($\rho\equiv\rho_0$). As the gas evolves in time, the density $\rho(\br,t)$ and the phase $\theta(\br,t)$ fluctuate due to the presence of the  disorder potential. 
We then exploit that the disorder potential is the smallest energy scale  to make use of perturbation theory. Precisely, we write 
$\rho(\br,t)=\rho_0+\delta \rho(\br,t)$ and linearize the Gross-Pitaevskii equation (\ref{eq:grosspitaevskii}) with respect to $\delta \rho(\br,t)$ and $\nabla\theta(\br,t)$ (without any assumption on the phase itself, which may be strongly fluctuating in this regime) \cite{Larre18}. This leads to the Bogoliubov-de Gennes equations
\begin{align}
&\dfrac{\partial \delta\rho}{\partial t}=-\dfrac{\rho_0}{m}\nabla^2\theta\\
&\dfrac{\partial \theta}{\partial t}=\dfrac{1}{4m\rho_0}\nabla^2\delta\rho-V(\br)-g\delta\rho,
\end{align}
which can be readily  diagonalized to provide the density,
\begin{align}
\delta\rho(\br,t)=-2\rho_0\int\frac{d^2\bk}{(2\pi)^2}\hat{V}(\bk)\frac{1-\cos(E_k t)}{\epsilon_k+2g\rho_0}e^{i\bk\cdot\br}
\label{rho_eq}
\end{align}
and the phase fluctuations,
\begin{align}
\theta(\br,t)=-\int\frac{d^2\bk}{(2\pi)^2}\hat{V}(\bk)\frac{\sin(E_k t)}{E_k}e^{i\bk\cdot\br}.
\label{theta_eq}
\end{align}
These expressions involve the well-known Bogoliubov dispersion relation, $E_k\equiv\sqrt{\epsilon_k(\epsilon_k+2g\rho_0)}$, where we recall that $\epsilon_k=\bk^2/2m$. The coherence function follows  from Eqs. (\ref{rho_eq}) and (\ref{theta_eq}) using standard procedures \cite{Castin03, Larre18, Martone18}:
\begin{align}
\label{eq_g1_largeg}
g_1(\br,t)& =\rho_0\exp\left\{-\int\frac{d^2\bk}{(2\pi)^2}B(\bk)(1-\cos\bk\cdot\br)\right.\nonumber\\
 &\left.
\times\left(\left[\frac{1-\cos(E_k t)}{\epsilon_k+2g\rho_0} \right]^2+
\left[\frac{\sin(E_k t)}{E_k} \right]^2\right)
 \right\},
\end{align}
where $B(\bk)\!\equiv\!\int d\br e^{i\bk\cdot \br}\overline{V(0)V(\br)}\!=\!2\pi V_0^2\sigma^2e^{-q^2\sigma^2/2}$ is the disorder power spectrum.
The physical content of Eq.  (\ref{eq_g1_largeg}) is the dynamical spreading of correlations in a Bose gas initially quenched from a weakly fluctuating state \cite{Cheneau12}, here in two dimensions. Unlike quench configurations more commonly encountered in the literature however \cite{Berges04, Cheneau12, Gring12, Trotzky12, Langen15, Martone18}, in our case the fluctuations are neither of quantum or thermal origin, but come from the random potential. 
In the present regime, scattering does not take place, so that the mean free time $\tau$ is no longer a relevant time scale. Instead, the short-time dynamics of the Bose gas is governed by the nonlinear time $\tau_\text{NL}\equiv 1/g\rho_0$. For  $t\gg\tau_\text{NL}$ and at the intermediate separations $\xi\ll r\equiv|\boldsymbol{r}|\ll 2c_s t$, where $\xi\equiv 1/\sqrt{g\rho_0m}$ is the healing length and $c_s=\sqrt{g\rho_0/m}$ is the speed of sound, Eq. (\ref{eq_g1_largeg}) simplifies to the time-independent algebraic law
\begin{equation}
g_1(\br,t)\simeq \rho_0\bigg[G(\sigma/\xi)\dfrac{\xi}{r}\bigg]^{\alpha}
\label{g1_algebraic}
\end{equation}
where $\alpha\equiv(V_0\sigma/\sqrt{2}g\rho_0\xi)^2$ and  $G(x)=\sqrt{2}x\exp\{[(6x^2+1)\exp(2x^2)E_1(2x^2)-3-\gamma]/2\}$, with $\gamma$ being the Euler-Mascheroni constant and $E_1$ the first-order exponential integral. A similar 2D algebraic scaling was found in \cite{Bardon-brun20}, in a configuration where the fluctuations were however encoded in the initial state and not in a disorder potential. Note that Eq. (\ref{g1_algebraic}) is both independent of time and of the precise shape of the disorder spectrum $B(\bk)$. It only depends on the small set of parameters $\{\xi,\sigma,g\rho_0, V_0\}$ and is thus, to a large extent, universal. Time independence and universality are characteristic properties of a pre-thermalization process, where a quenched system quickly converges to a fixed, thermal-like  point, from where it only departs very slowly \cite{Berges04, Gring12, Trotzky12, Langen15}. In the present scenario, the existence of a pre-thermalization regime requires the disorder amplitude to be very weak, so that the system is nearly integrable at short time. 
In the pre-thermal regime, the correlation function (\ref{g1_algebraic}) looks like the one of a uniform 2D Bose superfluid  \textit{at equilibrium and finite temperature} \cite{PetrovThesis}, and the gas exhibits an algebraically-decaying coherence \cite{Castin03}, in strong contrast with the multiple scattering regime (compare with Eq. (\ref{g1_smallg_theory})). 
Eq. (\ref{g1_algebraic}) breaks down at $r=2c_s t$, the boundary of a light cone within which correlations can spread. Out of the light cone, $g_1(\br,t)\simeq\rho_0[G(\sigma/\xi)\xi/4c_st]^\alpha$ reaches a time-dependent plateau, reminiscent of the perfect coherence of the initial plane-wave state.
Let us mention that to observe the algebraic law (\ref{g1_algebraic}), it is required to use of a \text{correlated} potential with $\sigma\leq \xi$ so to select the low-$k$ phonon modes $E_k\simeq c_s k$ in Eq. (\ref{eq_g1_largeg}). In particular, for an uncorrelated potential we have found no numerical evidence of algebraic decay in the coherence function.

\begin{figure}
\onefigure[scale=0.43]{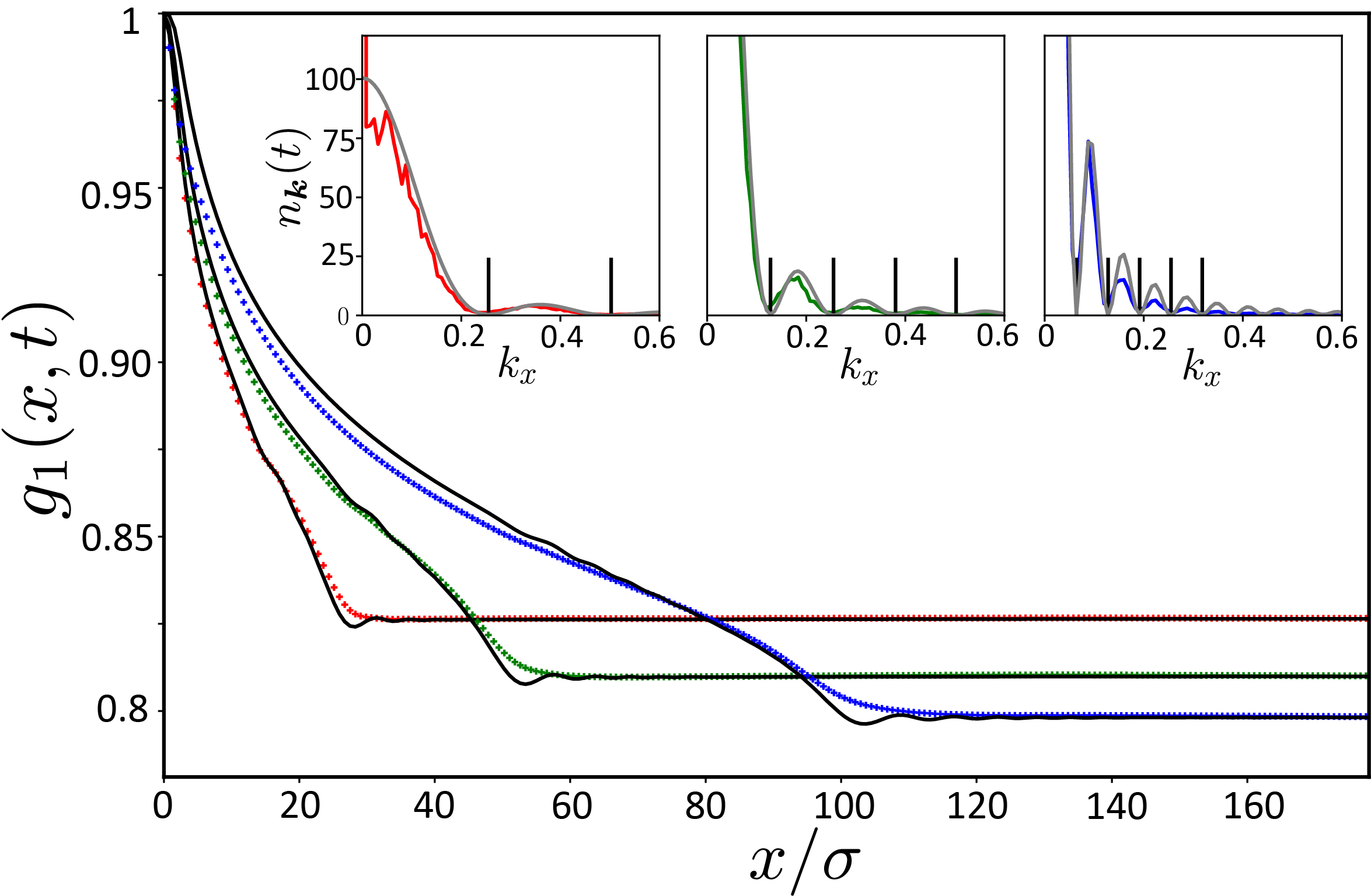}
\caption{
\label{fig_g1_largeg}
Cut along $x$ of the coherence function $g_1(\br,t)$ in the superfluid regime. 
Symbols show the numerical results obtained  at times $t=15\tau_\text{NL}$, $30\tau_\text{NL}$ and $60\tau_\text{NL}$ from bottom to top. Solid curves are the theoretical prediction, Eq. (\ref{eq_g1_largeg}), including a phenomenological $\beta(t)/k$ correction accounting for phonon collisions, as discussed in the main text. Fit values are $\beta(15\tau_\text{NL})=-0.0055$, $\beta(30\tau_\text{NL})=-0.094$ and $\beta(60\tau_\text{NL})=-0.152$. Insets show cuts along $k_x$ of the numerical momentum distributions at the same times (corresponding to Figs. \ref{fig_1}d-1-3), together with the theory, Eq. (\ref{profile_nk_largeg}).
Here $g\rho_0=1.5$, $V_0=0.4$, $k_0=0$, and the chosen system size is $L=250\pi$. With these parameters, $\tau_\text{NL}=0.67$ and $\sigma/\xi=1.22$.
}
\end{figure}

Within the linearization approach that led us to Eq. (\ref{eq_g1_largeg}), the disorder fluctuations are described in terms of independent Bogoliubov quasi-particles, which reduce to phonons at low-energy. Collisions between these phonons make the system slowly depart from the pre-thermal regime described by Eq. (\ref{g1_algebraic}) \cite{Buchhold15}. While the effect of these collisions remains small at short time, it was found in \cite{Bardon-brun20} that even at short time it is necessary to take them into account in order to achieve a quantitative agreement with numerical simulations for $g_1$. Collisions slowly change the momentum distribution of the quasi-particles \cite{Buchhold15}. We account for this change by adding a phenomenological, fitting parameter $\beta(t)/k$  in the second line of Eq. (\ref{eq_g1_largeg}), which becomes $(\beta(t)/k+[(1\!-\!\cos(E_k t))/(\epsilon_k\!+\!2g\rho_0)]^2+[\sin(E_k t)/E_k]^2)$. 
The factor $1/k$ is here introduced somewhat arbitrarily to reduce the weight of phonon collisions at short scale. We have verified that an alternative fitting option, independent of $k$, also allows us to   reproduce the numerical results \cite{Bardon-brun20}, albeit less accurately at short scale.
A comparison between Eq. (\ref{eq_g1_largeg}), modified according to this procedure, and the exact simulations for $g_1$, is shown in Fig. \ref{fig_g1_largeg} (parameters are here the same as those  in Figs. \ref{fig_1}d).
The agreement is excellent, except for the small spatial variations of $g_1$ in the vicinity of the light-cone boundary, which are not present in the simulation results. One reason might be a smoothing of these variations due to particle collisions.

When $V_0/g\rho_0\ll1$, one can expand the exponential in Eq. (\ref{eq_g1_largeg}), so that the momentum distribution $n_\bk(t)$ at $\bk\ne0$ is in first approximation given by the simple law
\begin{align}
n_\bk(t)\simeq \rho_0 B(\bk)
\left(\left[\frac{1\!-\!\cos(E_k t)}{\epsilon_k+2g\rho_0} \right]^2\!+\!
\left[\frac{\sin(E_k t)}{E_k} \right]^2\right).
\label{profile_nk_largeg}
\end{align}
Such a profile consists of concentric rings whose minima are located at positions $k_n=(\sqrt{2}/\xi)\{-1+[1+(\pi n/g\rho_0 t)^2]^{1/2}\}^{1/2}$, where $n$ is a non-zero positive integer. These rings are well visible in the distributions of Figs. \ref{fig_1}d-1-3. Physically, they originate from both the interference  between phonons (interference pattern $\sim\cos(2E_kt)$) and between the phonons and the mean field ($\sim\cos(E_kt)$). The spacing $\sim \pi/(c_s t)$ between the rings decreases in time, signaling that the interfering particles are further and further apart as time grows. 
Equation (\ref{profile_nk_largeg}) is displayed in the inset of Fig. \ref{fig_g1_largeg}, together with cuts along $k_x$ of the distributions in  Figs. \ref{fig_1}d-1-3 extracted from simulations. The agreement is again very good, in particular for the positions of the minima, indicated by vertical lines.

\section{Conclusion}

We have theoretically described the dynamics of a 2D, weakly interacting Bose gas in the presence of a random potential, and have identified two qualitatively different non-equilibrium regimes depending on the relative strengths of disorder and interactions. For interactions weaker than the disorder, the physics is that of multiple scattering, with interactions slowly thermalizing the energy distribution and a short-range coherence at short time. When interactions are stronger than the disorder, on the other hand, we recover a low-energy-physics regime: scattering ceases, the gas becomes a non-equilibrium superfluid, with  long-range correlations spreading within a light cone.

\acknowledgments
NC thanks Giovanni Martone for discussions. NC and DD acknowledge  financial support from the Agence Nationale de la Recherche (grants ANR-19-CE30-0028-01 CONFOCAL and ANR-18-CE30-0017 MANYLOK).


\begin{thebibliography}{99}
\bibitem{footnote1}
In the multiple scattering regime, the $g_1$ function has a small imaginary part at short time $t\lesssim\tau$ where the momentum distribution is not yet fully isotropic. In this manuscript however, we focus on times $t\gg\tau$, so that this imaginary part is negligible.
\end{thebibliography}
\end{document}